\documentclass[journal]{IEEEtran}
\usepackage{cite}
\usepackage{siunitx}
\usepackage{amsmath,amssymb,amsfonts}
\usepackage{algorithm}
\usepackage{algorithmic}
\usepackage{soul}
\usepackage{booktabs}
\usepackage{graphicx}
\usepackage{textcomp}
\usepackage{tabularx}
\usepackage{xcolor}
\usepackage[table]{xcolor}
\usepackage[normalem]{ulem}
\usepackage{multirow}
\usepackage{colortbl}
\usepackage{array}
\usepackage{makecell}
\newcolumntype{C}{>{\centering\arraybackslash}X} 
\begin{document}

\title{Plug-n-Play Three Pulse  Twin Field QKD\\
}

\author{Anagha~Gayathri,
        Aryan~Bhardwaj,
        Nilesh~Sharma,
        Tarun~Goel,
        Y.~V.~Subba~Rao,
        and~Anil~Prabhakar%

\thanks{A. Gayathri, N. Sharma, A. Bhardwaj, and A. Prabhakar are with the Centre for Quantum Information, Communication and Computing (CQuICC), Indian Institute of Technology Madras (IIT Madras), Chennai, India. A. Gayathri, N. Sharma, and A. Prabhakar are also with the Department of Electrical Engineering, IIT Madras. A. Bhardwaj is also affiliated with the Department of Physics, Indian Institute of Science Education and Research (IISER), Tirupati, India (email: anaghagayathri@smail.iitm.ac.in, nileshlns@smail.iitm.ac.in, aryanbhardwaj@students.iisertirupati.ac.in, anil@ee.iitm.ac.in).}

\thanks{T. Goel and Y. V. Subba Rao are with the Central Research Laboratory, Bharat Electronics Ltd., Bangalore, India (email: tarungoel@bel.co.in, yvsubbarao@bel.co.in).}%
}

\maketitle

\begin{abstract}
We present the experimental implementation of a three-time-bin phase-encoded Twin-Field Quantum Key Distribution (TF-QKD) protocol using a Sagnac-based star-topology plug-and-play architecture. The proposed encoding method leverages the relative phases of three consecutive time bins to encode two bits per signal. The Sagnac loop configuration enables self-compensation for both phase and polarisation drifts, eliminating the need for active stabilisation. However, field deployments are subject to rapid phase fluctuations caused by external vibrations, which can degrade interference visibility. We used the first time bin for real-time phase-fluctuation monitoring. Although this monitoring reduces the effective key generation rate, the system achieved a secure key rate of approximately $1.5 \times 10^{-5}$ bits per pulse, with a corresponding visibility of up to 87\% over a 50 km asymmetric optical fibre channel. These results demonstrate the practicality, stability, and scalability of the proposed three-time-bin TF-QKD protocol for real-world quantum communication networks.
\end{abstract}

\begin{IEEEkeywords}
Quantum key distribution (QKD), Measurement device independent  (MDI), Twin field QKD (TF), Sagnac interferometer, Quantum bit error rate (QBER), Security
\end{IEEEkeywords}

\section{Introduction}

Quantum Key Distribution (QKD) offers an unconditionally secure method for communication between remote users, with its security grounded in the principles of quantum mechanics rather than the computational assumptions underlying classical cryptographic algorithms. The field has undergone significant advancements since the first QKD protocol was introduced in 1984~ \cite{Bennett_2014}. While QKD protocols are theoretically proven to be unconditionally secure under the assumption of fully characterised devices \cite{shor2000simple, renner2008security, koashi2009simple}, practical implementations face substantial challenges. In practice, achieving complete device characterisation is extremely difficult, if not impossible. This gap opens potential vulnerabilities, creating side channels that eavesdroppers may exploit and, consequently, compromise the security guarantees of QKD systems \cite{portmann2022security}. The most critical of these side-channel threats arises from the non-ideal performance of single-photon detectors (SPDs) \cite{Lo14}. Several notable detector side-channel attacks have been reported, including detector blinding attacks, fake-state attacks, time-shift attacks \cite{makarov2009controlling, makarov2005faked, qi2005time}, and backflash attacks \cite{singh2025backflash}.

The detector-side vulnerabilities are mitigated by Measurement Device Independent QKD (MDI-QKD) protocol \cite{Lo12}, where the measurement operation is delegated to a third party, Charlie, who may be untrusted or act as an eavesdropper. Notably, the security of the protocol remains intact even when Charlie has access to the measurement outcomes. Experimental implementations of MDI-QKD using polarisation and time bin encoding have been reported \cite{polarizationbased,polarization2}. More recently, a differential phase-encoded MDI-QKD protocol has been presented \cite{Ranu21}, where information is encoded in the phase difference between consecutive bins in a three-time-bin superposition state. Despite these advancements, conventional QKD protocols, including MDI-QKD, are fundamentally limited in their achievable key rates over long distances without quantum repeaters or trusted nodes \cite{repeaterlessbound,repeaterlessbound2}. Twin-Field QKD (TF-QKD) overcomes this limitation and presents a promising approach for long-distance QKD \cite{TFQKD}. TF-QKD retains the architectural benefits of MDI-QKD and inherits its robustness against detector side-channel attacks. The first experimental realisation of TF-QKD was reported in 2019 \cite{TFQKD_EXP1}.

 Key challenges in implementing Twin-Field QKD include achieving optical phase alignment between Alice and Bob, and compensating for phase and polarisation fluctuations introduced by optical fibres. While active stabilisation techniques have been employed to mitigate these issues \cite{phasestab2}, they introduce additional system complexity and cost.
TF-QKD has also been demonstrated using a Sagnac interferometric architecture, which inherently provides self-compensation for phase fluctuations \cite{Sagnac_TFQKD}. Experimental studies have explored the performance of TF-QKD in Sagnac configurations over asymmetric optical fibre channels\cite{Assymetric_sagnac}, and a multi-user TF-QKD scheme employing time-multiplexing within the Sagnac architecture has also been proposed \cite{timemultiplexing}. Furthermore, the impact of Rayleigh backscattering in long-distance Sagnac-based TF-QKD systems has been analysed, and methods for mitigating its effects have been introduced \cite{longfiber}.  Nevertheless, these systems typically require passive polarisation compensation to counteract fibre-induced fluctuations.
Recently, a Sagnac-based TF-QKD network adopting a star topology was demonstrated \cite{star}, where polarisation fluctuations were compensated intrinsically by the network design. This implementation employed the sending-or-not-sending protocol, along with phase post-compensation techniques.

In this paper, we present the implementation of a three-time-bin-based TF-QKD protocol configured in a star topology. The proposed protocol utilises the first time bin to monitor rapid phase fluctuations, while the second and third bins are used for data encoding. We successfully implemented TF-QKD over a 50 km asymmetric fibre channel without employing any active phase or polarisation stabilisation techniques, or relying on post-processing-based compensation mechanisms. Table \ref{tab:tfqkd-comparison} benchmarks our protocol against recent TF-QKD implementations, highlighting its practical advantages and relevance to scalable quantum networking. The table includes the works that are directly relevant to and aligned with the approach adopted in this study.

\begin{table*}[!hbt]
\centering
\caption{Comparison of different TF-QKD implementations. Secure key rates are reported in bits per pulse. Channel length indicates the fibre spool length, and the value in parentheses represents total channel loss, including any additional optical attenuation.}
\label{tab:tfqkd-comparison}
\begin{tabular}{llllllll}
\toprule
\textbf{Reference} & \textbf{Implementation} & \textbf{Architecture} & \textbf{Year} & \makecell{\textbf{Channel length}\\\textbf{(km)}} & \makecell{\textbf{Secure}\\\textbf{keyrate}} & \makecell{\textbf{Compensation}\\\textbf{technique}} \\
\midrule
Zhong et al. \cite{Sagnac_TFQKD} & Phase-based & Sagnac (Ring)  & 2019 & 10 (44.9 dB) & $2.406 \times 10^{-5}$ & Passive polarisation \\
Zhong et al. \cite{timemultiplexing} & CAL19 Protocol & Sagnac (Ring)  & 2022 & 10 (51.16 dB) & $1.87 \times 10^{-6}$ & Passive polarisation \\
Mandil et al. \cite{longfiber} & NA & Sagnac (Ring) & 2025 & 200 & NA & Passive polarisation \\
Park et al. \cite{star} & SNS Protocol & Sagnac (Star)  & 2022 & 50 & $1.31 \times 10^{-4}$ & Phase post selection \\
Nascimento et al. \cite{nascimento2024passive} & NA & Sagnac (Star) & 2025 & 22 & NA & Inherent (star topology) \\
This work & Three-time-bin & Sagnac (Star) & 2025 & 50 & $1.5 \times 10^{-5}$ & Inherent (star topology) \\
\bottomrule
\end{tabular}
\end{table*}


\section{Protocol Description}

\begin{figure}[!ht]
    \centering
    \includegraphics[width=0.4\linewidth]{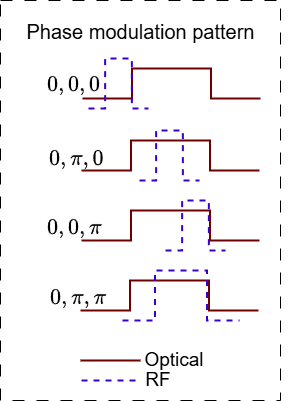}
    \caption{Phase encodings achieved by synchronising the PM's RF signal with the optical pulse envelope.}
    \label{fig:1}
\end{figure}

The proposed Twin-Field QKD protocol employs a three-time-bin superposition scheme to realise phase encoding using weak coherent states, as described in \cite{Diferentailphaseencoding}. A continuous-wave laser source is first modulated by an intensity modulator (IM) to generate optical pulses with an on-time of 3T, corresponding to three consecutive time bins ( T being the duration of one time bin). This pulse train is sent through a phase modulator (PM), which is driven by a random number generator to encode information in the relative phases between the time bins (Fig. \ref{fig:1}).

In this scheme, two binary bits are encoded per signal by modulating the optical signal randomly. Precise synchronisation between the RF signals driving the IM and PM ensures accurate alignment between the optical and electronic domains.

\begin{figure}[!ht]
    \centering
    \includegraphics[width=0.7\linewidth]{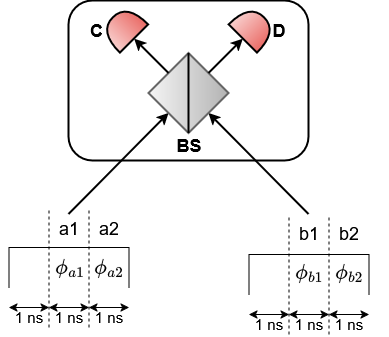}
    \caption{Interference of signals from Alice and Bob at Charlie. BS: Beam splitter, C: constructive port, D: destructive port. \(\phi_{a1}, \phi_{a2}, \phi_{b1}, \phi_{b2}\) represent phases encoded by Alice and Bob in time bins 2 and 3, corresponding to bits \(a1, a2, b1, b2\), respectively.}
    \label{fig:2}
\end{figure}

The protocol consists of the following steps:

\begin{itemize}
    \item \textbf{Step 1:} Alice and Bob independently prepare weak coherent pulses using the three-time-bin encoding method. The information is encoded in the relative phases between the time bins, where no encoding is applied to the first time bin. (i.e., it serves as a phase reference).
    
    \item \textbf{Step 2:} The encoded states are transmitted to an untrusted intermediate node, Charlie, via optical fibre.
    
    \item \textbf{Step 3:} At Charlie, the signals from Alice and Bob interfere at a 50:50 beam splitter. The two output ports are monitored by single-photon avalanche diodes (SPADs), labelled C (constructive port) and D (destructive port), as illustrated in Fig.~\ref{fig:2}.
    
    \item \textbf{Step 4:} Charlie announces the measurement outcomes over a public channel, including the detection time bin and the detector port (constructive or destructive).
    
    \item \textbf{Step 5:} Based on the publicly announced information, Alice and Bob determine the correlation between their encoded bits and perform key sifting and distillation.
\end{itemize}

Table~\ref{tab:decode-table} outlines the possible detection events at Charlie and the corresponding bit relationships. In this scheme, detections in the constructive output port during the second and third time bins are denoted as $C_2$ and $C_3$ respectively, while $D_2$ and $D_3$ refer to detections in the destructive output port at the same time bins. Random bits \(a_1, b_1\) are encoded by Alice and Bob in the second time bin, and \(a_2, b_2\) are the bits encoded in the third time bin. The first time bin is unmodulated by Alice and Bob and is expected to yield constructive interference. Photon detections in the first bin are not used for key generation but for monitoring the phase drift.

\begin{table}[!ht]
\centering
\caption{Detection outcomes and corresponding bit conclusions at Charlie.}
\label{tab:decode-table}
\begin{tabular}{cccccc}
\toprule

\textbf{S.no.} & \multicolumn{4}{c}{\textbf{Outcomes}} & \textbf{Conclusion} \\

 & \textbf{$C_2$} & \textbf{$C_3$} & \textbf{$D_2$} & \textbf{$D_3$} & \\
\midrule
1 & 1 & 0 & 0 & 0 & \( a1 = b1 \) \\

2 & 0 & 1 & 0 & 0 & \( a2 = b2 \) \\

3 & 0 & 0 & 1 & 0 & \( a1 \oplus b1 = 1 \) \\

4 & 0 & 0 & 0 & 1 & \( a2 \oplus b2 = 1 \) \\
\bottomrule
\end{tabular}
\end{table}

\section {Experimental Description}
\begin{figure}[!ht]
   \centering
   \includegraphics[width=1\linewidth]{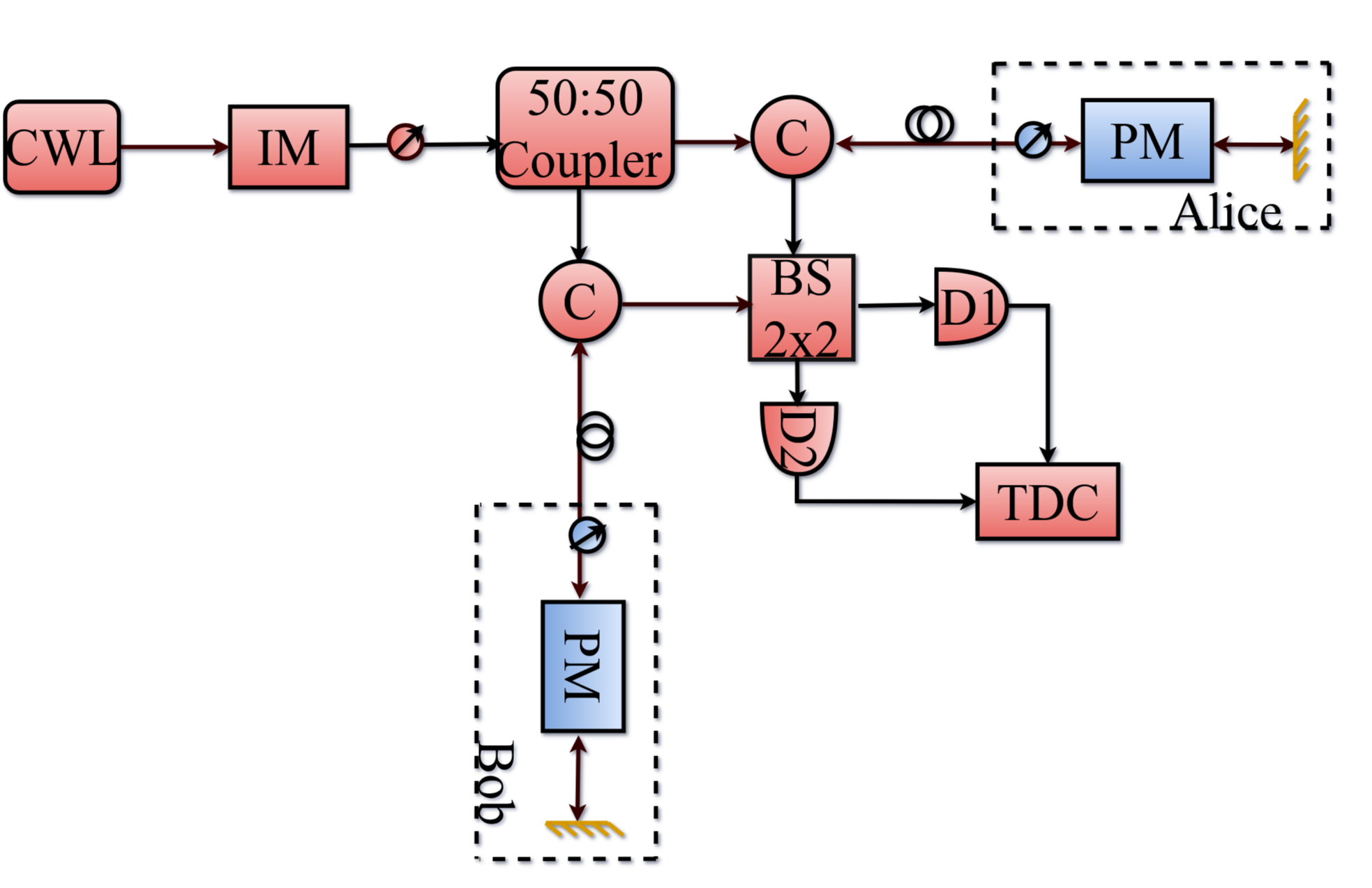}
  \caption{Schematic of PnP twin field QKD setup with phase encoding.} 
  \label{fig:3}
 \end{figure}

We have implemented the TF-QKD protocol using a plug-and-play (PnP) architecture. As illustrated in Fig. \ref{fig:3}, the central node, Charlie, houses a continuous-wave laser source, which is modulated into optical pulses using an intensity modulator (IM). These pulses are then split by a beam splitter and sent to the remote users, Alice and Bob, via separate optical paths.

Upon receiving the pulses, Alice and Bob encode binary information into the relative phase of the incoming signals using phase modulators. The modulated pulses are subsequently attenuated to weak coherent states and reflected toward Charlie along the same optical paths. Thus, ensuring the compensation of polarisation fluctuations. Moreover, since the optical pulses are sourced from a single laser at Charlie, the architecture inherently obviates the need for phase locking between independent lasers at Alice and Bob.

At Charlie's station, the returning pulses from Alice and Bob are interfered with at a beam splitter. Two SPADs are placed at the constructive and destructive output ports of the beam splitter to detect interference outcomes. The detection events, i.e., SPAD clicks, correspond to the relative phase encodings applied by Alice and Bob, enabling the generation of correlated bits for key distillation.

The RF drive signals for the IM, PM, and SPAD gates are generated by a field-programmable gate array (FPGA), with programmable timing control to a resolution of 62.5 \unit{\pico\second}. The RF signals driving the IM and SPADs are configured with a 3 \unit{\nano\second} ON duration and a 32 \unit{\nano\second} period, while the PM is driven with a one \unit{\nano\second} ON duration, maintaining the same period.

A key advantage of the PnP configuration is its intrinsic polarisation self-compensation. This is realised through a Faraday rotator, which rotates the polarisation state of the incident light by 90$^\circ$ upon reflection. Since the optical signal traverses the same fibre path from Charlie to Alice and back, any polarisation fluctuations encountered during the forward path are automatically compensated upon return, maintaining alignment at the beam splitter.

Despite polarisation compensation, the system remains sensitive to random phase fluctuations in the fibre, which adversely affect the interference output. The interference visibility (discussed in a later section) at Charlie serves as the primary metric for evaluating the system's overall performance.  Fig.~\ref{fig:5} illustrates that the visibility fluctuates significantly over time, ranging from 0.7 to 0.16, due to these phase instabilities.

To counter such phase noise, active phase stabilisation methods have been proposed and experimentally demonstrated \cite{phasestab2}. While effective, these methods often add considerable complexity and cost to the system. Alternatively, the Sagnac loop architecture has gained attention in the literature owing to its intrinsic phase self-compensation. For our PnP system, we modify and implement a Sagnac loop-based approach, combining the benefits of passive phase stability with the robustness of the PnP configuration.

\begin{figure*}[htbp]
\centering
\includegraphics[width=0.9\linewidth]{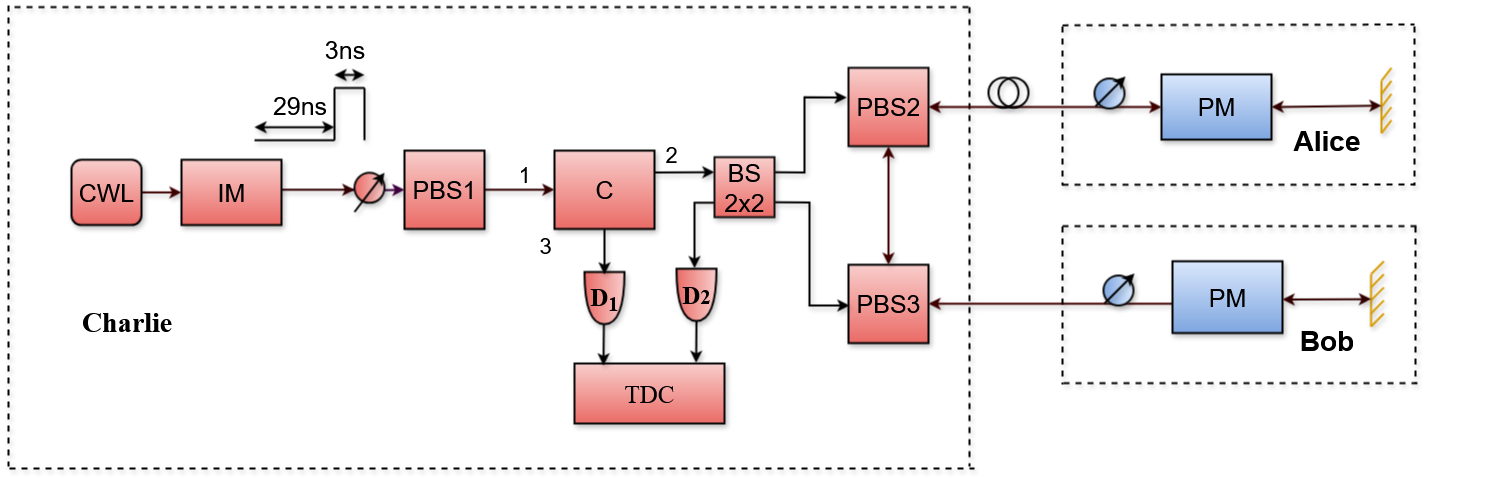}
\caption{Schematic of modified Sagnac-based Twin field QKD setup }
\label{fig:4}
\end{figure*}

In this modified architecture, shown in Fig.~\ref{fig:4}, self-phase compensation is achieved due to the common-path nature of the setup. Optical pulses originating from Charlie first pass through a polarising beam splitter (PBS1), which transmits only one polarisation component (say horizontal) toward Alice and Bob via a beam splitter.

At Alice's (or Bob's) station, the incoming pulse passes through a phase modulator and is reflected by a Faraday mirror (FM), which rotates its polarisation state by 90$^\circ$ before sending the pulse back toward Charlie. This polarisation rotation ensures that the returning signal is reflected by the PBS2 (PBS3) at Charlie and directed to Bob (Alice).

At Bob's (or Alice's) station, the pulse is phase modulated and reflected by the Faraday mirror, which again rotates its polarisation. Note that this polarisation rotation makes it horizontal again. On this return path to Charlie, PBS3 (or PBS2) allows the pulse with the rotated polarisation to reach the beam splitter. It is important to emphasise that the pulse is modulated only once when it passes through the phase modulator for the second time.

At Charlie, the returning pulses, which have been phase-encoded by Alice and Bob, interfere at the beam splitter. The resulting interference pattern is determined by the relative phase shifts applied by the two parties, enabling secure key generation based on the detection outcomes.

\subsection{Comparison}
In this section, we compare a standard TF-QKD configuration (Fig.~\ref{fig:3}) with the modified TF-QKD scheme using a star-topology PnP configuration (Fig.~\ref{fig:4}). Visibility quantifies the contrast of the interference pattern and is defined as,

\begin{equation}
V = \frac{P_{c} - P_{e}}{P_{c} + P_{e}},
\end{equation}

\noindent where $P_c$ and $P_e$ represent the photon counts corresponding to correct and erroneous detection events, respectively.

\begin{figure}[!ht]
\centering
\includegraphics[width=1\linewidth]{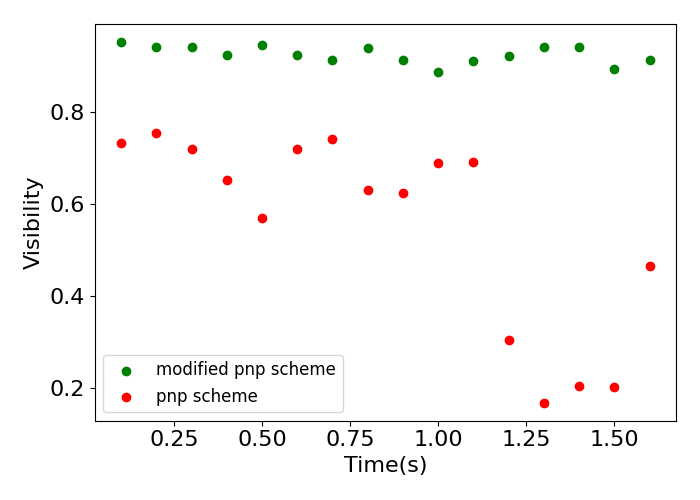}
\caption{Interference visibility at Charlie when Alice and Bob send a 0-0-$\pi$, for two different TF-QKD configurations.}
\label{fig:5}
\end{figure}

To measure visibility, we applied a known encoding pattern: 0-0-$\pi$ at Bob, while no modulation was applied at Alice. Under this configuration, constructive interference is expected in the first and second time bins, while the third bin should ideally exhibit no photon counts. Thus, visibility in this case is calculated as:

\begin{equation}
V = \frac{(n_1 + n_2) - n_3}{n_1 + n_2 + n_3},
\end{equation}

\noindent where $n_1$, $n_2$, and $n_3$ are the photon counts in the first, second, and third time bins, respectively. The calculated visibility over time is plotted in Fig.~\ref{fig:5}.

As shown in Fig.~\ref{fig:5}, the modified star-topology PnP scheme (green trace) demonstrates a clear improvement in visibility compared to the standard TF-QKD configuration (red trace). This improvement indicates that random phase fluctuations are more effectively compensated in the modified setup.

This enhancement can be attributed to the common-path nature of the star-topology Sagnac-inspired architecture. This design inherently creates symmetry in the optical paths and minimises relative phase misalignment, which in turn effectively stabilises the interference, even though the absolute phase of the system remains random.

We further assessed the robustness of the system by varying the channel length between Alice and Charlie using fibre spools. The average visibility over a 5 \unit{\second} window was recorded for different fibre lengths, as shown in Table \ref{tab:vis-table}.

\begin{table}[!ht]
\centering
\caption{Average interference visibility at Charlie for different fibre channel lengths.}
\label{tab:vis-table}
\begin{tabular}{cc}
\toprule

\textbf{Distance (km)}  & \textbf{Visibility(\%)} \\

\midrule
0 & 90 \\

10  & 89.7 \\

20 & 88 \\

50  & 87.8 \\
\bottomrule
\end{tabular}
\end{table}

To fully exploit the self-phase compensation nature of the proposed scheme, the phase fluctuations experienced by the clockwise and counter-clockwise propagating optical signals in the Sagnac loop must remain effectively identical. This condition holds when the phase fluctuation rate is relatively low, typically within a few kilohertz for a 50 km fibre channel.

However, in field-level deployments, external mechanical disturbances can induce rapid phase fluctuations that exceed the compensation bandwidth of the system. Such disturbances lead to degraded interference visibility, resulting in a higher quantum bit error rate (QBER) and ultimately necessitating key discarding.

To address this challenge, the proposed three-time-bin TF-QKD scheme allows phase monitoring using the detection events in the first time bin. These detections can serve as a reference to estimate the phase drift. Furthermore, a post-processing algorithm based on the visibility of the first time bin detections can be employed to recover valid key bits, even under dynamic phase fluctuations.
\section{Results and discussion}
To evaluate the performance of the modified Sagnac scheme (Fig.~\ref{fig:4}), we performed a proof-of-principle experiment involving all three parties. An optical signal generated at Charlie was sent to Alice and Bob, where it was modulated according to the protocol, and then returned to Charlie for interference and detection as explained in the earlier section. This setup allows us to analyse the resulting QBER and the corresponding generation of secure keys.

\subsection{Proof of principle experiment} 

Phase encoding at Bob is achieved by introducing programmable delays to the RF signal driving the phase modulator. This shifts its modulation relative to the optical pulse arrival time. The optical signals from Alice and Bob are interfered at Charlie using a beam splitter, and the interference output is detected using a SPAD operated at an efficiency of approximately 10\%. The SPAD gate timing is carefully aligned to maximise photon detection probability.

\begin{figure}[!ht]
\centering
\includegraphics[width=\linewidth]{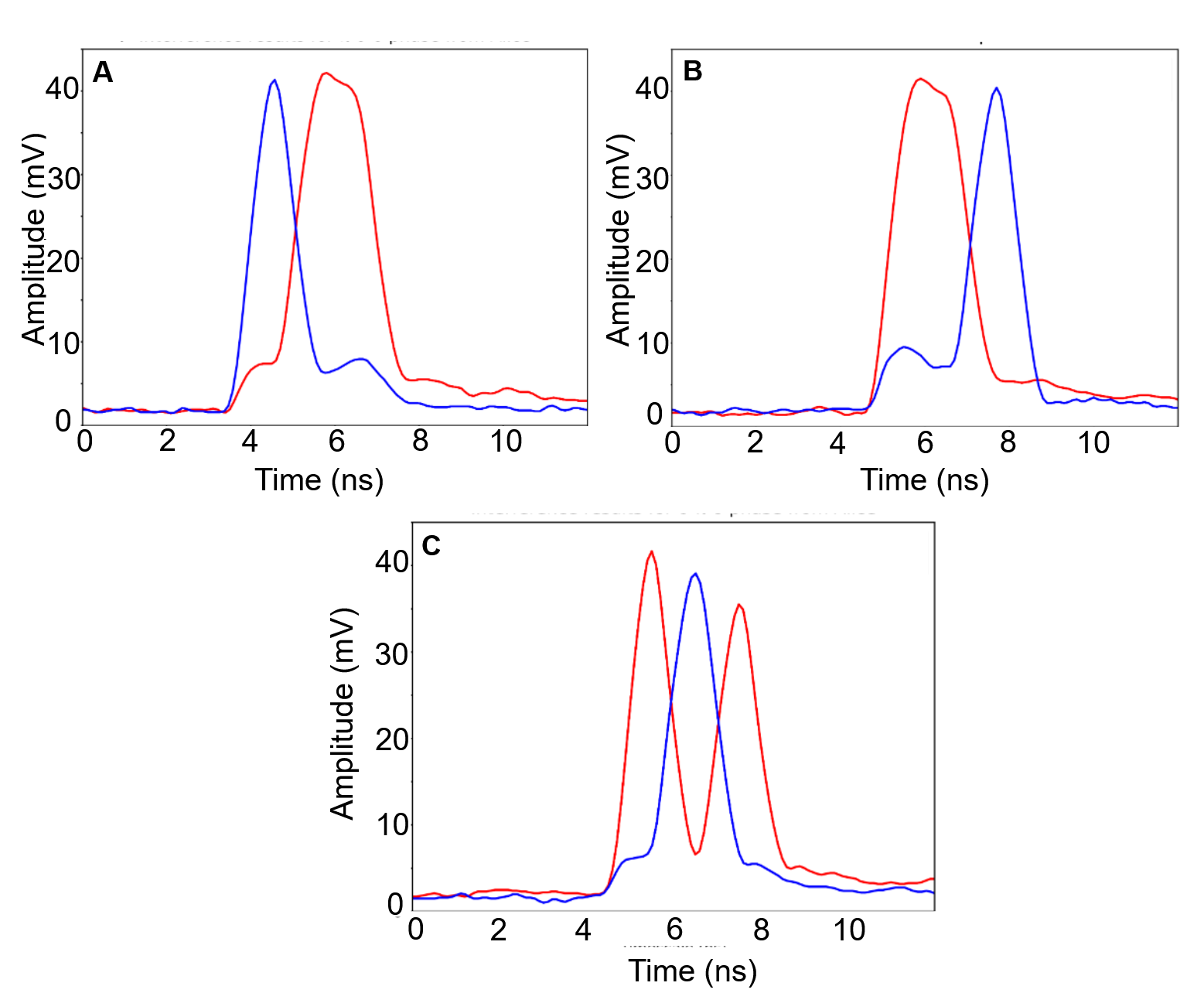}
\caption{Classical interference results for different modulation patterns applied at Bob, with Alice returning unmodulated pulses. (A) Bob applies a $\pi$ phase in the first time bin. (B) Bob applies a $\pi$ phase in the third time bin. (C) Bob applies a $\pi$ phase in the second time bin. Constructive and destructive port outputs are shown in red and blue, respectively.}
\label{fig:8}
\end{figure}

Before applying attenuation to achieve the desired mean photon number, we observed the classical interference output on a digital storage oscilloscope (DSO) using a 5~GHz bandwidth avalanche photodiode. Fig.~\ref{fig:8} shows the output measured by Charlie for various modulation patterns applied. The red traces represent the output at the constructive port, and the blue traces correspond to the destructive port.

Following this, the photodiodes are replaced by single-photon detectors, and a time-to-digital converter (TDC) is used to collect photon detection timestamps. In Figs.~\ref {fig:8}B and \ref{fig:8}C, constructive interference is observed in the first time bin, indicating no relative phase misalignment between the optical pulses from Alice and Bob. In contrast, Fig.~\ref{fig:8}A shows destructive interference in the first time bin, corresponding to an intentional relative $\pi$ phase shift applied at Bob. It is important to note that such a phase shift can arise from fast phase fluctuations induced by external disturbances along the optical channel.

To address this, we implemented a post-processing algorithm that is capable of identifying and correcting phase misalignments. Specifically, the algorithm monitors photon counts in the first time bin over a defined time interval. A decrease in counts below a predetermined threshold indicates a potential $\pi$ phase misalignment. During such intervals, the algorithm flips the bits associated with the second and third time bins to recover the correct logical encoding. This correction is applied prior to QBER analysis and key distillation. It is important to note that the current algorithm is designed to detect and correct primarily near $\pi$ phase misalignments; however, the approach can be extended in the future to accommodate arbitrary phase shifts for enhanced robustness.

\subsection{ QBER calculation and Temporal filtering}

The QBER quantifies the error rate in a QKD system. It is calculated by comparing the received key bits with the transmitted bits and determining the ratio of incorrect bits to the total number of received bits. A high QBER may indicate the presence of an eavesdropper or system-level noise; in such cases, the corresponding keys are discarded to ensure security.

One of the primary challenges in \hyphenation{time-bin superposition-based}encoding lies in the speed limitations of electronic components operating in the picosecond regime. Errors can arise from various sources, including timing jitter in components such as SPADs and TDCs, as well as transient responses in RF amplifiers. To mitigate these issues, guard bands are introduced at the edges of each time bin, allowing tolerance against such fluctuations \cite{Diferentailphaseencoding}.

While increasing the guard band reduces QBER, it also decreases the sifted key rate, as fewer photon events fall within the valid detection window. Therefore, optimising the guard band involves balancing this trade-off by maximising the secure key rate, which depends on both QBER and the sifted key rate.
Secure key rate ($R$) is derived for the scheme as, 
\begin{equation}\label{eq:skr}
R = R_{\text{sift}} [1 - h(e_b) - h(e_p)],
\end{equation}
where $R_\text{{sift}}=4r$ is the sifted key rate with $r$ being the sifting probability due to one of the four concluding and mutually exclusive events, $e_b$ is observed QBER and $h(x)=-x\text{log}_2(x)-(1-x)\text{log}_2(1-x)$ is the Shannon entropy. Information leakage to an eavesdropper is estimated in terms of phase error rate $e_p$, which is upper bounded using $e_b$ \cite{Sharma24}. The factor four is due to four different useful detections for sifting in the scheme.  

\begin{figure}[!ht]
         \centering
         \includegraphics[width=1\linewidth]{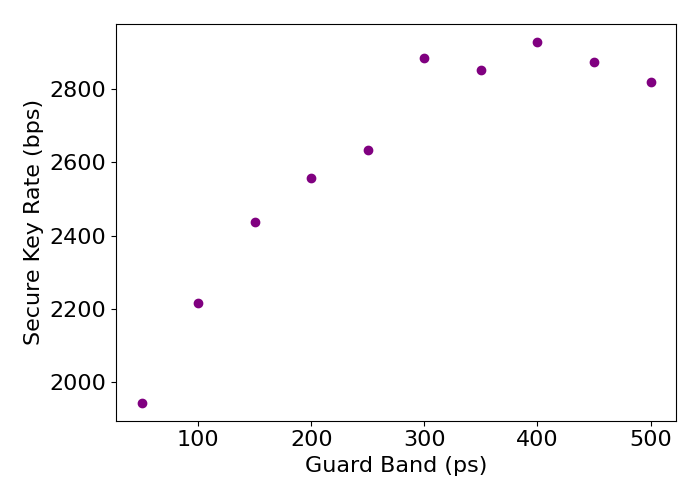}
         \caption{Observed secure key rate for different guard bands.}
         \label{fig:9}
    \end{figure}
We evaluated the sifted key rate and average QBER across different encoding patterns for various guard band widths. The resulting secure key rates are plotted in Fig.~\ref{fig:9}. The graph shows that the secure key rate increases with the guard band and saturates around 300 ps, indicating that a 300 \unit{\pico\second} guard band provides an optimal trade-off between error suppression and key throughput. To further examine the system performance, we measured photon detection statistics at Charlie for fixed phase encoding patterns, and the QBER was estimated using the timestamps from the constructive port.

To assess the system's robustness under different transmission conditions, we conducted the experiment using fibre spools of varying lengths: 10 km, 20 km, and 50 km, placed between Alice and Charlie. At Bob's side, a short patch cord was used. Average QBER over a 1-second interval was calculated for each configuration. 

\subsection{Back scattering effects}
One of the major disadvantages of the PnP configuration is the effect of backscattered signals. These signals can arise from Rayleigh, Raman, or Brillouin scattering processes. However,  Raman and Brillouin scattering are negligible due to their higher power thresholds and frequency-shift characteristics, making Rayleigh scattering the dominant backscattering mechanism \cite{ subacius2005backscattering}. Note that the PnP architecture, which launches the classical pulses from the receiver, results in Rayleigh backscattering of these classical signals that temporally overlap with the quantum signal, thereby significantly increasing the QBER.

We modelled the probability of detection due to backscattering of the classical signals sent by Charlie to Alice (and Bob) for channel length $l$ between Alice and Bob, as \cite{peng2008multi, subacius2005backscattering},
\begin{equation}\label{eq:bksctrng}
    P_B(l)=2(1-\eta)N\overline{\mu}\beta t_{\text{ON}}\eta_{\text{det}}.
\end{equation}
Here $\eta=10^{-\alpha l/10}$ is the transmittance of the channel, $\alpha$ is the attenuation coefficient of the fibre in dB/km, $N$ is the repetition rate, $\overline{\mu}$ is the mean photon number per pulse sent to Alice (Bob), $\beta$ is the backscattering coefficient, $t_{\text{ON}}$ is the gate on time for the SPADs and $\eta_{\text{det}}$ is the detection efficiency of the SPADs. The factor two accounts for the backscattering in both paths from Charlie to Alice and to Bob. The simulation parameters are shown in Table \ref{tab:param}.

\begin{figure}[!ht]
         \centering
         \includegraphics[width=1\linewidth]{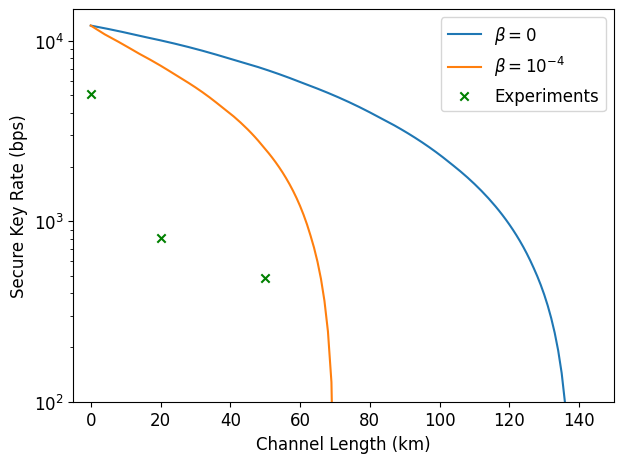}
         \caption{Estimated Secure key rate (SKR) from experiments against expected SKR from simulation over different channel lengths. Solid lines represent simulated SKR for different values of scattering coefficient $\beta$.} 
         \label{fig:11}
    \end{figure}

We then simulated the performance of the PnP TF-QKD using the models discussed in Eq. \ref{eq:skr} and \ref{eq:bksctrng} and ref. \cite{Sharma24}, considering the source trusted. The output power from Charlie is -55dBm at 1550 nm. wavelength. The corresponding detection events introduce errors with a probability of 0.5. The detections due to backscattering are analogous to dark count events, in the sense that they also contribute errors with the same probability. Therefore, in our simulations, we effectively increased the dark count probability by the estimated probability of backscattered photon detections. The resulting QBER values thus incorporate the impact of backscattering. The simulated results were subsequently compared with the experimental results in Fig. \ref {fig:11}. 

A key practical consideration is the use of a 10$\%$ efficient SPAD in the experimental setup, which results in a lower secure key rate compared to systems employing high-performance detectors \cite{star}. Nevertheless, the system maintains a positive secure key rate using a more economical and commercially viable detector. The results also confirm the system's capability to operate with asymmetric channel lengths, a valuable feature for practical network deployment.
\begin{table}[h]
    \centering
    \caption{Simulation parameter. \\$N$: repetition rate, $\overline{\mu}$: mean photon number per transmitted pulse, $\beta$: backscattering coefficient, $t_{\text{on}}$: gate on time, $\eta_{\text{det}}$: detection efficiency of the SPADs, $\alpha$: attenuation coefficient of the fibre.}
    \label{tab:param}
    \begin{tabular}{c c c c c c}
    \toprule
     $N$ & $\overline{\mu}$&$ \beta$ & $t_{\text{ON}}~ \text{(in \unit{\nano\second})}$ &$\eta_{\text{det}}$ & $\alpha ~\left(\text{in~} \unit{\decibel}/\unit{\kilo\meter} \right)$\\
     \midrule
      $31.25 \text{~M}$   & 40 & $10^{-4}$ & 3   & 0.1 & 0.2\\
         \bottomrule
    \end{tabular}
    
\end{table}

\section {Conclusion}

In this work, we have experimentally demonstrated a three-time-bin phase-encoded TF-QKD protocol implemented within a Sagnac-based star-topology PnP architecture. This common-path configuration inherently provides the phase compensation, significantly improving system stability and performance. Our comparative analysis with a conventional TF-QKD configuration demonstrated a substantial improvement in interference visibility, achieving a maximum of 87$\%$ over a 50 km optical fibre channel, validating the robustness and practical suitability of the architecture for quantum communication networks. 

The proof-of-principle was conducted over an asymmetric channel configuration with a 50 km fibre link from Alice to Charlie and a short patch cord from Bob to Charlie without active phase or polarisation control. This demonstrates the system's compatibility with asymmetric channel lengths, enabling PnP operation, which is especially important for practical network deployments.


The efficacy of self-phase compensation, however, is contingent on the phase fluctuation rate within the optical fibre. It remains effective only when the round-trip propagation time through the channel (Charlie → Alice → Bob → Charlie) is shorter than the characteristic timescale of phase drift. In real-world deployments, mechanical disturbances can induce rapid phase shifts, degrading interference visibility and increasing the QBER. To evaluate the system's performance against such disturbances, we simulated an intentional $\pi$  phase shift. In this test scenario, our protocol utilises the first-time bin as a phase reference to monitor and estimate phase fluctuations. We further implemented a post-processing algorithm that detects and corrects the phase misalignments by analysing photon detection statistics in the first time bin, thus preserving key integrity under fluctuating channel conditions.

Distinguishing itself from conventional ring topologies, the star topology inherently facilitates polarisation compensation, which in turn permits scalable network deployment. Furthermore, this architecture allows seamless integration of additional users through orthogonal polarisation states, eliminating the need for direct optical links between end users. Leveraging these scalability advantages, we envision evolving this platform into a fully networked QKD system. In this model, a central node distributes phase-stable quantum signals to multiple users, facilitating simultaneous and secure key generation across a multi-user quantum network.

\section*{Acknowledgments}
This work was partly supported by the Mphasis F1 Foundation. The authors also acknowledge the financial support provided by Bharat Electronics Limited (BEL), India. AG, AB, and NS would like to thank Ashutosh Singh and Valliamai Ramanathan for their valuable inputs on the manuscript.
\bibliography{ref}
\end{document}